# Time-based analysis of the NBA hot hand fallacy

Samuel Henry

Abstract

The debate surrounding the hot hand in the NBA has been ongoing for many years. However, much of the existing work on this theme has focused on only the very next sequential shot attempt, often on very select players. This work looks in more detail the effect of a made or missed shot on the next series of shots over a two-year span, with time between shots shown to be a critical factor in the analysis. Also, multi-year streakiness is analyzed, and all indications are that players cannot really sustain their good (or bad) fortune from year to year.

## I. Introduction

The debate surrounding the hot hand effect (or fallacy) [1-3] has been ongoing for decades, but lately has received more attention. In a recent article in the New York Times, it was implied once again that the hot hand is real, reigniting the debate [4]. And more recently it was shown that repetition in the case of NBA free throws can influence percentages [5]; and it was also shown to exist in Major League Baseball [6].

Over the years, many articles have been written with the intent of proving the hot hand in the NBA. Typically, when trying to prove a theory or hypothesis, the default assumption is null, or that the theory is false. In the case of field goals, there has been no clear evidence for the hot hand effect, at least in any widely accepted way. It has even been shown that there exists a counter effect to the hot hand [7]. As will be confirmed in this work, the next shot a shooter takes is *less* likely to go in than if he made the shot.

Most of the aforementioned work has focused on only the very next sequential shot (as opposed to series of several shots), also regardless of time elapsed in between shots. This work looks in more detail on the effect of a made or missed shot on the next *sequence* of shots, with time between shots also shown as a critical factor in the analysis.

## II. Autocorrelation methodology

By definition, an autocorrelation is a cross-correlation of a signal with itself. When a time series signal is auto-correlated, you start with two identical time series data sets and shift one by some number (*shot lag*). Then multiply each element in one signal with the analogous element in the other (that was just shifted), then sum [8].

And finally, divide the final result by a normalization term. In this case, the *signal* is just a sequence of ones (for made shots) and zeros (missed shots), with appropriate filters applied. Figure 1 shows the autocorrelagram of all shot sequences for all players from 2014-2016 averaged for all players over both seasons. The dataset was extracted from play-by-play data on basketball-reference.com.

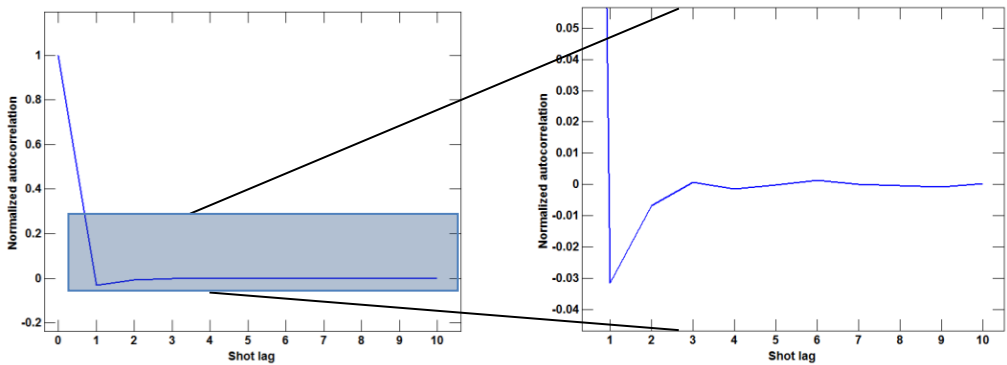

Figure 1: Average autocorrelegram for all players for sequences within a single game; data from 2014-2016

Essentially, each point on the blue curve above represents the correlation between any given shot (either made or missed), with the $n^{th}$ subsequent attempt. The vertical-axis is closely related to the correlation (often denoted R), as if we had two independent variables *x* and *y*. For reference, with shot lag of 0, each point is 100% correlated (R = 1) because the comparison is made basically between two identical data sets (*x* is equal to *y*). In the case of shot lag equal to one, *x* would be the original sequence of made or missed shots, and *y* is the sequence "next" shots. Would these two variables be correlated? Figure 1 shows that with a shot lag of one, these are actually negatively-correlated. That is for every make, the player is slightly more likely to miss.

One can easily notice this in the curve, where it dips sharply from one to below zero for lags of one and two. For a truly random signal that is sufficiently long, the correlation should drop from unity to zero in the first lag, and all subsequent lags. But even after at a lag of 4-5, and again 7-9 there is some slight negative correlation. In no subsequent shot in Figure 1 is the result strongly positive.

3. Impact of time between shots

Next, let's look at a very similar plot to Figure 1, except with a minute's filter applied. For example, the blue color in Figure 2 filters sequences where all shots were taken at most 1 minute from one another (in terms of game time). One can see the autocorrelegram has large dependence on time between shots. The more time that passes, the more the hot hand "counter-effect" is diminished. Next time you see someone make a shot, keep a look out for a quick subsequent shot and take note of how many times that shot is a make (but don't forget to count the misses).

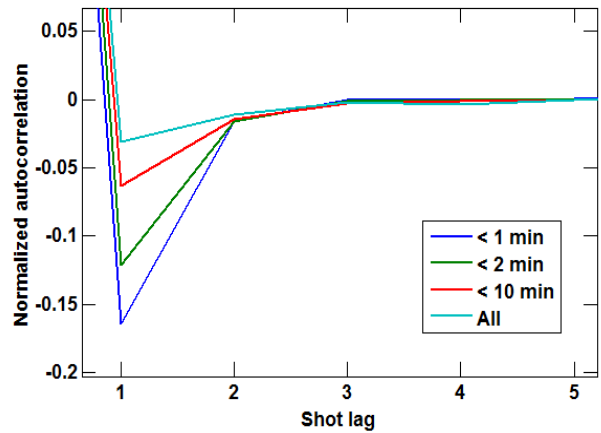

Figure 2: Autocorrelegram for all players with different colors filtering various time between shots

It essentially shows that if a player has the opportunity to shoot quickly after a make, he will be more likely to miss (relative to the same shot attempt with more time elapsed). But another natural question arises: how much quicker is the player going to take a next shot after a make, versus a miss?

Next, let's look how likely a made shot will result in a quicker shot attempt. Figure 3 plots the average FG% ($y$-axis) of a given shot plotted vs. time ($x$-axis) between two adjacent shots, rounded to nearest minute. This plot was generated by putting the entire 2-year sample into buckets of average time between shots. For example, sequences with an average delta between shots of 2 minute indicates the made FG% of the *original* shot was 41%. Conversely, if you took 8 minutes to take your next shot, then your average FG% for the original shot was 33%, before the 8 minutes lapses. A filter for only shots greater than 15 feet was applied, where the hot hand argument is presumably most relevant.

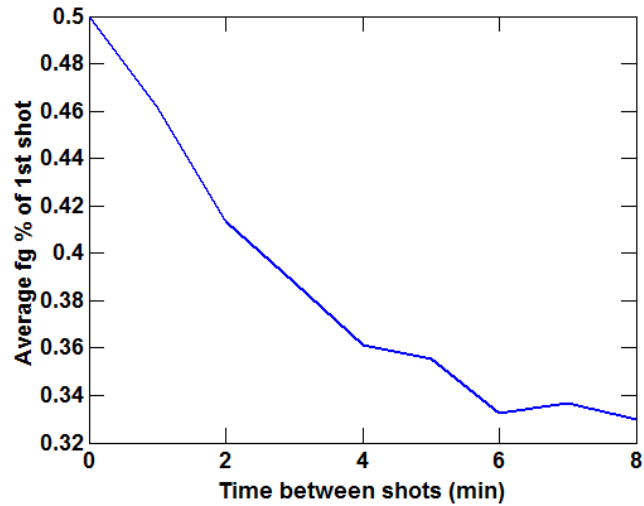

Figure 3: Average FG % of 1st shot plotted vs. time until next shot (with a filter of greater than 15 feet applied)

In essence, Figure 3 shows that the FG% is anti-correlated with the time passing until the next shot. This is an interesting finding and in a general sense agrees with the previous plot showing the various minute filters. The interpretation is basically the same: after a made shot, a player seems to "force" the next shot.

4. Year-year correlation of streakiness

Often one of the counter arguments to the hot hand fallacy is something like: "Sure, most players don't on average have a hot hand, but surely some players in the NBA can have a statistically significant hot hand?" Or put another way some players with itchy trigger fingers after a make are clouding the data and "hiding" disciplined players that indeed display the hot hand.

First, it should be noted, that every year there are a few examples of statistically significant hot hands. But this is a core tenant of the hot hand fallacy. Statistically anomalous streaks do exist by nature, a small percentage of the time, and it is up to the human to properly calibrate what streaks are completely random, and what streaks are indicative of something (i.e., can be used to predict *something* in the future).

Take for example Sasha Vujacic. See his personal autocorrelagram shown below in Figure 4, using data from 2013-2016. This correlgram used data from within the same game, and shots taken beyond 15 feet. Both home (blue line) and away (red line) show he was a better shooter after make– and again four shots after a make (but only while at home); holding over a three year span! Perhaps this is more than simple coincidence!

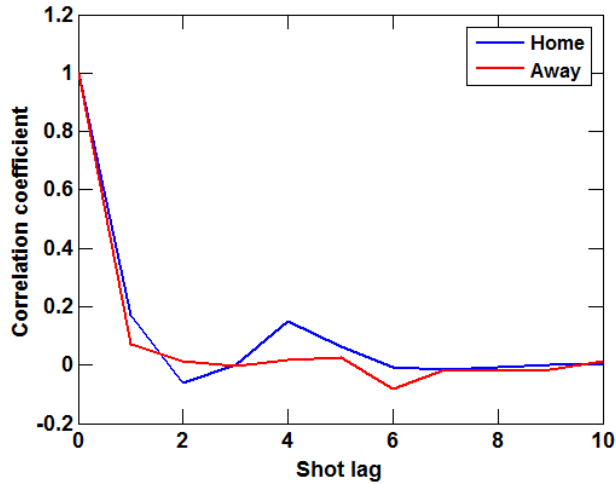

Figure 4: Sasha Vujacic autocorrelagram from 2014-2016 showing a potential to have the hot hand during this time

Cherry picked examples such as this can be seen or heard often. But as it turns out, every year there will be players that has an extended hot streak (or for a set of three years there will still be some, albeit fewer). But does that mean it's sustainable, or have any predictive power for the future? In other words, wouldn't the disciplined shooters stay disciplined (hot hand) and the non-disciplined shooters stay non-disciplined (cold-hand)?

Figure 5 below attempts to answer the question of weather can the streakiness of any given player be used to predict future streakiness. Plotted below are the same induvial data points used to create Figure 1, except *without* averaging over players or years. That is, each point represents the average correlation between made/misses in a shot sequence for a player in either 2014 (x-axis), or 2015 (y-axis). Lag of 1 (meaning consecutive shot attempts) and lag 2 were chosen here.

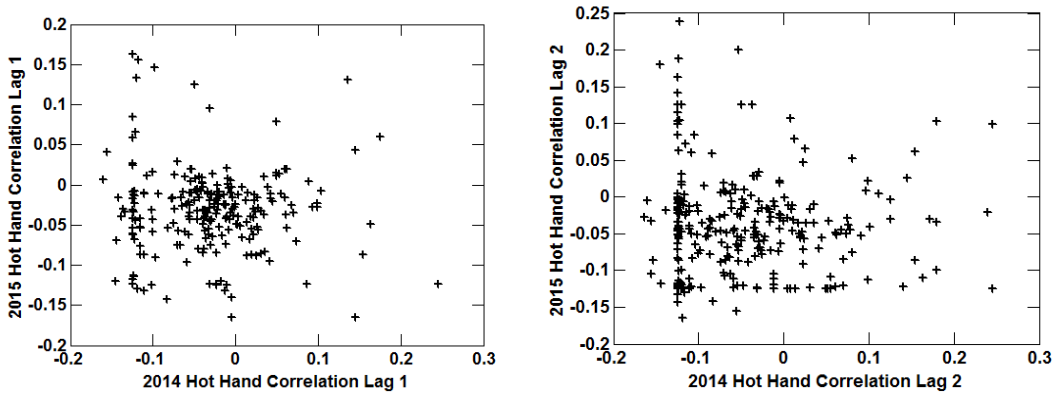

Figure 5: Each point represents the average correlation between made/misses in a shot sequence for a player in an entire year. *x*-axis is 2014, *y*-axis is 2015. Left plot shows a lag of 1 (correlation of consecutive shots); right shows lag of 2

One can see from Figure 5 that there is very little correlation from year to year in terms of streakiness. Let's just say we have a non-disciplined player that is "anti-streaky" (i.e., after a make likely to miss), and some others that are positively streaky. There is simply no evidence that streakiness for these respective players will continue. This can be considered evidence to further the hot hand fallacy in basketball.

## 5. Conclusion

So far, in this work it was shown that not only does the hot hand not exist, but after a made shot, the negative effects can last up to 2-3 shots in the future. This is particularly true when the time until the next shot is short. Also, it was shown that the FG% of any given shot is inversely related to the time until that player takes the next shot. Lastly, there was a test to see if there are some disciplined shooters could consistently be "streaky." This could obscure the overall hot hand data. But streakiness of any given player was shown to have no correlation year-year, negating the argument that certain players have prolonged streakiness.

It should be noted that there are two ways to look at the hot hand fallacy. One is that the offensive player *forces* a worse next shot. The other side that is that the defensive player pays a little extra attention on the subsequent shot attempt. But if you think about it these are really saying the same thing. If the defense pays extra attention on the subsequent shot, and the player shoots anyway, then it is a *worse* shot. The takeaways are fairly obvious. While the proposed defensive strategy is a bit more complicated, on offence it seems clear that players should simply be aware of this fact and act accordingly (i.e., don't take a bad shot just because you made the last one).

## Citations